\documentclass[reprint, amsmath,amssymb, aps, superscriptaddress, floatfix]{revtex4-1}
\usepackage{graphicx}
\usepackage{dcolumn}
\usepackage{bm}
\usepackage{natbib}

\usepackage{amsbsy,amssymb,latexsym,amsfonts,amsmath}
\usepackage{graphicx,color}

\begin{document}
\bibliographystyle{plainnat}

\preprint{APS/123-QED}
\title{Nature of chiral phase transition in two-flavor QCD}

\author{K.-I. Ishikawa}
\affiliation{Graduate School of Science, Hiroshima University,Higashi-Hiroshima, Hiroshima 739-8526, Japan}

\author{Y. Iwasaki}
\affiliation{Center for Computational Sciences, University of Tsukuba,Tsukuba, Ibaraki 305-8577, Japan}

\author{Yu Nakayama}
\affiliation{Kavli Institute for the Physics and Mathematics of the Universe (WPI), Todai Institutes for Advanced Study,  Kashiwa, Chiba 277-8583, Japan}
\affiliation{Department of Physics, Rikkyo University, Toshima, Tokyo 177-8501, Japan}

\author{T. Yoshie}
\affiliation{Center for Computational Sciences, University of Tsukuba,Tsukuba, Ibaraki 305-8577, Japan}

\date{\today}

\begin{abstract}
	
We investigate the nature of the chiral phase transition in the massless two-flavor QCD using the renormalization group improved gauge action and the Wilson quark action on $32^3\times 16$, $24^3\times 12$, and $16^3\times 8$ lattices. 
We calculate the spacial and temporal propagators  of the iso-triplet mesons in the pseudo-scalar ($PS$),  scalar ($S$), vector ($V$) and axial-vector ($AV$) channels on the  lattice of three sizes.
We first verify that the RG scaling  is excellently satisfied for all cases.
 This is consistent with the claim that the chiral phase transition is second order.
Then we compare the spacial and temporal  effective masses  between the axial partners, i.e. $PS$ vs $S$ and $V$ vs $AV$, on each  of the three size lattices. We find the effective masses of all of six cases for the axial partners agree remarkably.  This is consistent with the claim that at least $Z_4$ subgroup of the $U_A(1)$ symmetry in addition to the $SU_A(2)$ symmetry is recovered at the chiral phase transition point.

\end{abstract}

\maketitle

\section{introduction}
Chiral phase transition in QCD plays a fundamental role in the history of the universe. In particular, it is crucial to understand when and how the phase transition occurred from the quark gluon state to the hadronic state.
However, it is a challenging problem since it depends on non-perturbative properties of QCD and even in the idearistic case 
of massless two degenerate flavor quarks, fundamental problems such as the order of the phase transition and the meson spectroscopy  at the phase transition point are under a  hot debate. See for example,~\citep{Pisarski:1983ms}\citep{Iwasaki:1996ya}\citep{AliKhan:2001ek}\citep{Ejiri:2009ac}\citep{Bazavov:2012qja}\citep{Aoki:2012yj}\citep{Buchoff:2013nra}\citep{Pelissetto:2013hqa}\citep{Nakayama:2014sba}\citep{Sato:2014axa}\citep{Kanazawa:2015xna}\citep{Brandt:2016daq}\citep{Tomiya:2016jwr}.

 In this article we aim at settling {the issue
of the chiral phase transition in the massless $N_f = 2$ flavor QCD}.
The QCD Lagrangian with $N_f=2$ massless quarks is invariant under $SU_L(2)\times SU_R(2) \times U_V(1) \times U_A(1)$ flavor rotations. 
The $SU_L(2)\times SU_R(2)$ symmetry that is spontaneously broken to $SU_V(2)$ (i.e. iso-symmetry) in the QCD vacuum at zero temperature is recovered above the chiral phase transition temperature  $T_c \sim \Lambda_{QCD}$.
On the other hand, the  $U_A(1)$ symmetry is broken by the quantum anomaly down to $Z_2$ in the vacuum. The fate of the $U_A(1)$ symmetry at finite temperature is what we would like to pursue in this article. Naively, we expect that it is gradually recovered as the temperature increases
and eventually is fully recovered in the $T \rightarrow \infty$ limit.

The question of whether the $U_A(1)$ symmetry is recovered or not near $T_c$ is of  
phenomenological as well as theoretical importance.
As  first discussed  by Pisarski and Wilczek \citep{Pisarski:1983ms}, the order of the phase transition may depend on the fate of the $U_A(1)$ symmetry.

In our previous work \citep{Ishikawa:2017hka}, we identified the chiral phase transition point by monitoring the number of iterations in HMC algorithm.
However we have not yet verified that the $SU_A(2)$ chiral symmetry is really recovered at the transition point.
One test of the chiral symmetry recovery is to compare the spectrum of vector ($V$) and axial-vector ($AV$) mesons.
Having verified the recovery of the $SU_A(2)$ chiral symmetry,
our main target is whether the $U_A(1)$ symmetry is recovered at $T_c.$
We directly compare the spectrum of the iso-triplet $U_A(1)$ partner; pseudo-scalar ($PS$) vs the scalar ($S$) mesons to see this.
If the $U_A(1)$ symmetry is recovered, not only the mass but also the propagators must agree between the $U_A(1)$ partners.


In our recent article \citep{Ishikawa:2017hka}  
we also derived the RG scaling relation for the effective masses of mesons in the $PS$ and $V$ channel at the chiral phase transition point  under the assumption of the second order phase transition.
 Our numerical {data of }simulations for $PS$ and $V$ mesons on the lattices of three sizes, i.e. $32^3\times 16$, $24^3\times 12$, and $16^3\times 8$ lattices, were excellently on the scaling curves.
This is  consistent with the second order phase transition.

In this article, we extend the RG scaling analysis to the cases of the $S$ and $AV$ mesons to investigate the fate of $U_A(1)$ at the chiral phase transition temperature. 
We  compare the spacial and temporal  effective masses  {between the $U_A(1)$ partners, i.e. $PS$ vs $S$, and between the $SU_A(2)$ partners, i.e. $V$ vs $AV$, on each  of three size lattices.} We find that the effective masses of all of six cases agree with axial partners, each on the lattice of three different sizes.
 This is consistent with the claim that $U_A(1)$ (in addition to $SU_A(2)$ symmetry) is recovered at the phase transition point. More precisely, what we will show is that at least $Z_4$ subgroup of the $U_A(1)$ symmetry is recovered.

The organization of the paper is as follows.  
After describing our setup in section 2, we revisit RG equations and derive RG scaling relations in section 3.
 In section $4A$ we give the job parameters, and in section $4B$ we locate the chiral phase transition points.
 In section $4C$ we show our numerical results for the spacial direction with the verification of the scaling relation and the recovery of the $U_A(1)$ symmetry and $SU_A(2)$ symmetry. We present the similar analysis for the temporal direction in $4D$.
  Finally summary and  discussions are given in section 5.

\section{Action and Observables}
\label{action}
We define continuous gauge theories as the continuum limit of lattice gauge theories,
defined on the Euclidean lattice of the size $N_x=N_y=N_z=N_s$ and $N_t$.  We impose an anti-periodic boundary condition in the temporal direction for fermion fields and periodic boundary conditions otherwise. We define the aspect ratio $r=N_s/N_t$, and keep it fixed when we change the lattice size.



In this article, we study $SU(3)$ gauge theories with degenerate $N_f=2$ Dirac fermions in the fundamental representation.
We employ {the RG improved gauge action and the Wilson quark action}~\citep{RG-improved}.
The theory is defined by two parameters; the bare coupling constant $g_0$ and the bare degenerate quark mass $m_0$ at ultraviolet (UV) cutoff.
We also use, instead of $g_0$ and $m_0$, 
$\beta={6}/{g_0^2}$
and the hopping parameter
$K= 1/2(m_0a+4)$. 

Two words about our choice of the lattice action are in order. 
 First of all, the reason why we use the RG improved gauge action is that the action is close to the renormalized trajectory~\citep{RG-improved} and the proper contributions from topological excitations are taken account~\citep{Itoh:1984pr}. These two aspects are essential to study the chiral phase transition  since the phase transition occurs at  strong coupling, and the the recovery of the symmetry may depend crucially on the effects of topological excitations.
 
 Secondly, in the recent literature it might be trendy to use more ``chiral symmetry friendly" fermion action such as domain wall fermions to investigate the chiral phase transition. We, instead, opt to use the Wilson fermion with careful fine-tuning. We already know that at zero temperature  the large mass difference between iso-singlet and non-singlet can be explained in the Wilson fermion formalism. Together with the  analyses of the chiral symmetry based on Takahashi-Ward identities~\citep{Bochicchio:1985xa}, there are, in principle, no fundamental theoretical and phenomenological problems to treat massless fermions in the Wilson fermion formalism. The advantage of the Wilson fermion is that we can increase the lattice size with less cost, which is important in taking the continuum as well as thermodynamic limit.

We measure the mass of hadrons such as the pseudo-scalar meson mass $m_{\text{PS}}$.
The quark mass $m_q$ is defined through Takahashi-Ward identities.
The main observables  we study in this article is the meson propagator.
The spatial propagator of a meson is defined by
\begin{align}
G_{s}(x) = \sum_{t,y,z} \langle \bar{\psi}\gamma_H \tau_a\psi(x,y,z,t) \bar{\psi}\gamma_H \tau_b\psi(0,0,0,0) \rangle,
\end{align}
where $H$  is the corresponding gammma matrix for the $PS$,  $S$, $V$ or $AV$ channel, and $\tau_a$ is the Pauli matrix in flavor space, which means that we study the propagator in the iso-triplet representation.
 We also study the temporal propagator defined in the similar way.

{
$Z_4$ subgroup of the $U_A(1)$ acts on the Dirac fermion by $\psi \to e^{\frac{i\pi}{2} \gamma_5} $ so that we have $\bar{\psi}\gamma_S\tau_a\psi \leftrightarrow \bar{\psi}\gamma_{PS}\tau_a\psi$  but leave $\bar{\psi}\gamma_V\tau_a \psi$ and  $\bar{\psi}\gamma_{AV}\tau_a \psi$ invariant. Therefore if the $Z_4$ symmetry is recovered, the propagator of the $S$ and $PS$ mesons must be the same. In order to study the full recovery of the $U_A(1)$ symmetry, we need to study the other correlation functions. Similarly, if the $Z_4$ subgroup of the $SU_A(2)$ symmetry is recovered, the propagator of the $V$ and $AV$ mesons must agree because it exchanges these two mesons.
}

\section{RG relation}
In our previous work, we have studied the consequence of the RG equation. Focusing on the massless quark trajectory, the RG equation in the vicinity of the critical point is given by
~\citep{DelDebbio:2010ze}\citep{DeGrand:2009mt}\citep{DelDebbio:2010hu}
\begin{eqnarray}G_s(n_s; g, N, \mu)_H &=& \nonumber \\
{\left(\frac{N'}{N}\right)}^{-2\gamma} &G_s& (n_s{'}; g{'}, N{'},  \mu{'} )_H\label{s-RG} \end{eqnarray}
for the spatial propagator with suffix $s$. The similar RG equation applies to the temporal propagator {with suffix $t$}.

Here $n_s=n_x$ and $\mu{'}= \mu/s $ and  $N_s{'}=N_s/s,$ $N_t{'}=N_t/s,$ and $n{'}=n/s$ with $s$ being the change of the scale under the renormalization.
The UV renormalization scale $\mu$ in lattice theories is set by the inverse lattice spacing $a^{-1}.$  
Note $N_s a=L_s$ and $N_t a =L_t$ are kept constant.
The relation between $g'$ and $g$ is determined by the RG beta function.

Our criterion for the chiral phase transition on the finite lattice is based on the ``on $T_c$ method" discussed in \citep{Ishikawa:2017hka}. We first determine the massless quark lines and then change $\beta$ to see if the inversion of the Dirac operator is possible or not{, keeping the molecular steps.} The location $g(N)$ where it becomes impossible depends on the lattice size and it is regarded as the chiral phase transition point. They are determined in 
\citep{Ishikawa:2017hka} and given in subsection 4B.

We define the scaled effective mass by
\begin{equation}\mathfrak{m}(n_t; g, N) = N \ln \frac{G(n_t; g, N)}{G(n_t+1; g, N)},\label{scaled effective mass}\end{equation}
suppressing $\mu$.
In the continuum limit $N \rightarrow \infty,$ 
we obtain the RG equation 
\begin{equation}\mathfrak{\check{m}}_s(\tau, g(N), N)=\mathfrak{\check{m}}_s(\tau, g(N{'}),  N{'}).\label{RG for s-mass}\end{equation}
Similarly we obtain the RG relation for the temporal effective mass
\begin{equation}\mathfrak{\check{m}}_t(\tau, g(N), N)=\mathfrak{\check{m}}_t(\tau, g(N{'}),  N{'}).\label{RG for t-mass}\end{equation}
Eqs.~(\ref{RG for s-mass})  and (\ref{RG for t-mass}) 
are key relations which are valid when the chiral phase transition is second order.

At the chiral phase transition points, one may solve the RG equation as
\begin{equation}
\mathfrak{\check{m}_{s}}(\tau, g(N), N)=\mathfrak{F_{s}(\tau)} , \label{universal}
\end{equation} 
and
\begin{equation}
\mathfrak{\check{m}_{t}}(\tau, g(N), N)=\mathfrak{F_{t}(\tau)} , \label{t_universal}
\end{equation} 
where $\mathfrak{F}_{s}(\tau)$ and  $\mathfrak{F}_{s}(\tau)$ are RG invariant. It means that the effective masses scale  with different lattice sizes on the chiral phase transition points.


\section{Numerical results}
\subsection{Job parameters}
We perform simulations with two degenerate quarks 
on $32^3 \times 16,$ $24^3 \times 12$  and $16^3 \times 8$ lattices  to investigate the scaling of the effective masses of mesons, and the $Z_4$ subgroup of the $U_A(1)$ symmetry from the mesons spectroscopy.
The algorithm we employ is the blocked HMC algorithm \citep{Hayakawa:2010gm}.
We choose the run-parameters in such a way that the acceptance of the HMC Metropolis test is about $70\%\sim 90\%.$
The statistics are 1,000 MD trajectories for thermalization and $1000\sim5000$ MD trajectories for the measurement.
We estimate the errors by the jack-knife method
with a bin size corresponding to 100 HMC trajectories. See Table 1.
 
{Here is a  cautious remark for the gauge configuration generation~\citep{Ishikawa:2013tua}\citep{Ishikawa:2015nox}. There are quasi-stable states characterized  by the spacial Polyakov loops. One has to choose a random initial configuration and take a reasonably wide step size in order to get configurations for the lowest energy state.}

\subsection{Chiral phase transition points}
The chiral phase transition points on the finite lattice are identified in
\citep{Ishikawa:2017hka}, as listed below.
\begin{itemize}
\item
$ \beta_{*}\simeq 2.8; K_{*}=0.1455$ on the $32^3\times16$ lattice;
\item
$ \beta_{*}\simeq 2.6; K_{*}=0.1480$ on the $24^3 \times 12 $ lattice;
\item
$ \beta_{*}\simeq 2.3; K_{*}=0.1547$ on the $16^3 \times 8 $ lattice.
\end{itemize}

The lattice spacing  is estimated $a\simeq 0.057$ fm at $\beta=2.8$ and  the lattice  size $L_s=32 \times a$ is $\sim 1.85$ fm.\citep{Ishikawa:2017hka}

\subsection{ Spacial propagators}
Now let us show numerical results of the spatial effective masses measured at the critical points. Our goal is to test the RG scaling relations for the $S$ and $AV$ channels. In particular we would like to see if they match the ones in $PS$ and $V$ channels to verify the recovery of the chiral symmetry.


First we show the data  in Fig.1 to verify the RG scaling relations.
{We note that the errors of data, here and hereafter, are smaller than the size of the marks, which is less than 1\%}.
We plot the scaled spatial effective masses defined in eq.(\ref{scaled effective mass}) in terms of $\tau=n_s/N_s$ 
to test the scaling relation (\ref{RG for s-mass}).
We overlay the data on the three lattices of $32^3\times 16$, $24^3\times 12$ and $16^3\times 8$.
We see that all the data of four channels (i.e. $PS$, $V$, $S$, and $AV$ channels) are excellently on the scaling curve except for two points at short distance ($n_s =1, 2$) on each of the lattices.

Next, we compare the effective masses of the axial partners, i.e. $PS$ vs $S$ and $V$ vs $AV$,
on each lattice of $32^3\times 16$, $24^3\times 12.$ and $16^3\times 8$.
The results given in Fig.2 show that the effective masses of the axial partners are  in remarkable agreement with each other. This is consistent with the claim that at least $Z_4$ subgroup of the $U_A(1)$ (in addition to the $Z_4$ subgroup of $SU_A(2)$) is recovered at the critical phase transition point.

\subsection{Temporal propagators}
We also measure the effective temporal  masses for four channels on the lattices of three sizes.
Since the data points are half of the spacial effective masses,
the RG scaling behavior is not as clearly seen, but the data are consistent with the RG scaling as shown in Fig. 3.
On the other hand, it is more than excepted to see the agreement of the 
 effective masses of the axial partners, i.e. $PS$ vs $S$ and $V$ vs $AV$,
on the each lattice of  $32^3\times 16$, $24^3\times 12$ and $16^3\times 8$ shown in Fig.4.
They agree with each other even on the smallest lattice of  $16^3\times 8$.
This also suggests that $Z_4$ subgroup of the $U_A(1)$ (in addition to the $Z_4$ subgroup of $SU_A(2)$) is recovered at the critical phase transition point.

 \section{summary and discussion}
There are mainly two approaches to investigate the fundamental problems such as the order of the phase transition and the symmetry  at the phase transition point in massless two-flavor QCD. One  is to analyze the finite temperature path integral directly in $d=1+3$ dimensions
    and the other is to use the $d=3$ dimensional effective field theory. 
 
Our approach is the former \citep{Ishikawa:2017hka}.
We have derived the RG scaling relations for the meson propagators under the assumption of the second order transition.
When the RG equation is evaluated in the vicinity of the UV fixed point $g_0=0$ and $m_0=0$,
the quark mass term is relevant and the gauge coupling is marginal.
Along the RG trajectory from the UV fixed point to the IR critical point,
the beta function does not possess a zero, that is,  the beta function is negative 
along the RG trajectory. The gauge coupling constant (or temperature) is relevant, and we have to tune it to obtain the criticality. 
Numerically, we have verified the RG scaling relations, which means that the number of relevant operators does not increase along the RG trajectory. In addition, we have verified the recovery of the chiral symmetry  by comparing the effective masses between the axial partners of the iso-triplet mesons at the phase transition temperature.

  The most straightforward interpretation of our results is that the chiral phase transition is second order and at the chiral phase transition temperature at least $Z_4$ subgroup of the $U_A(1)$ symmetry is recovered. 

  With this respect, there is a theoretical analysis of the recovery of the $U_A(1)$ symmetry in $d=1+3$ dimension based on the structure of the Dirac eigenvalue distributions above the chiral phase transition temperature \citep{Aoki:2012yj}\citep{Kanazawa:2015xna}\citep{Tomiya:2016jwr}\citep{Cohen:1996ng}\citep{Lee:1996zy}\citep{Evans:1996wf}. The recent papers \citep{Aoki:2012yj}\citep{Kanazawa:2015xna} in particular show that when the thermal distribution of the Dirac eigenvalues are sufficiently analytic around zero above the chiral phase transition temperature, {at least} $Z_4$ subgroup of the $U_A(1)$ must be recovered, in agreement with our results.

Let us now discuss the  analyses in the $d=3$ dimensional effective field theory. 
  First, Pisarski and Wilczek studied the problem in the framework of the perturbative  $\epsilon$  expansion of the three-dimensional Landau-Ginzburg-Wilson model.
 They calculated the RG beta functions at the lowest order and concluded that the order of phase transition depends on the fate of the $U_A(1)$ symmetry at the chiral phase transition.
 If the breaking of the $U_A(1)$ symmetry is large, there is an IR fixed point with $SU_L(2)\times SU_R(2)$ ($= O(4)$) symmetry and the phase transition must be the second order, while if the breaking vanishes,
 there is no IR fixed point and therefore the phase transition is first order. Our results at first sight contradict with their claim.
 
 One should note, however, that their analysis may not be trusted in the $\epsilon=1$ limit (i.e. $d=3$), where the validity of the $\epsilon$ expansion is under question, and the later RG analysis seems to suggest the  opposite to what Pisarski and Wilczek claimed.
 Indeed higher order corrections to the RG beta functions have been studied   with delicate resummation \citep{Pelissetto:2013hqa}  and they concluded that there exists an IR fixed point with the $SU_L(2)\times SU_R(2) \times U_A(1)$ symmetry. Therefore, if we believe in the existence of such a fixed point in the $d=3$ Landau-Ginzburg-Wilson model, the second order phase transition with the recovery of the $U_A(1)$ symmetry is possible.
 Another novel approach in favor of such a fixed point comes from the conformal bootstrap~\citep{Nakayama:2014sba}\citep{Nakayama:2016jhq}. The predicted critical exponents are all in agreement with each other.

Thus these theoretical analyses and numerical results in $d=1+3$ dimensions suggest that the chiral phase transition is second order and the $U_A(1)$ symmetry or at least the $Z_4$ symmetry  is recovered  at the chiral phase transition temperature.
On the other hand, the approach in $d=3$ dimensions suggests the chiral phase transition is second order and  the $U_A(1)$ symmetry is recovered at the chiral phase transition temperature. 
A remaining issue is which symmetry of the $U_A(1)$ symmetry or the $Z_4$ symmetry is actually realized.

In this connection, the recent analysis of the conformal bootstrap \citep{Nakayama:2016jhq} in $d=3$ dimensional Landau-Ginzburg-Wilson model suggests that the 
 above mentioned RG fixed point with the $SU_L(2)\times SU_R(2) \times U_A(1)$ symmetry has a relevant operator that is invariant under the $Z_4$ symmetry. This means that  if only the $Z_4$ symmetry is recovered,
 one cannot reach the  $SU_L(2)\times SU_R(2) \times U_A(1)$ symmetric RG fixed point without fine-tuning.
  Therefore one may conclude that this RG fixed point cannot explain the second order chiral phase transition if the symmetry is no larger than $Z_4$. It is in conflict with what we observed in the scaling behavior without an extra fine-tuning than the temperature.
 

Thus it is most plausible that the $U_A(1)$ symmetry is recovered at the critical point
under the basic assumption that the fixed point {so far found} in the $d=3$ dimensions corresponds to the chiral phase transition in massless $N_f=2$ flavor QCD. 
It is desirable to study higher point functions in lattice simulations to directly test
whether the recovered symmetry is $Z_4$ or $U(1)$.
 
 Finally, there is a further possibility that the phase transition is actually the first order, but the RG flow is accidentally slow so that one cannot distinguish it from the second order phase transition with the lattice size studied so far. Such possibilities are suggested by \citep{Sato:2014axa}. 
  To test this scenario numerically, we need to perform simulations 
 on lattices with larger aspect ratios $r$  toward the thermodynamic limit.

We would also like to thank S. Hashimoto for useful discussion.
The calculations were performed on Hitachi SR16000 at KEK under its Large-Scale Simulation Program and HA-PACS computer at CCS, University of Tsukuba under HA-PACS Project for advanced interdisciplinary computational sciences by exa-scale computing technology.

\begin{table*}
\caption{Job parameters and spectroscopy at $K_c$ 
on the $32^3\times 16$,  $24^3\times 12$ and $16^3\times 8$ lattices.
The masses  $m_{PS}$, $m_{V}$, $m_{S}$ and $m_{AV}$  are the value at the largest $n$, although they are  decreasing as $n$ increases.}
\begin{tabular}{lccccccccc}
\hline
size & beta & $K$ & sweeps &acc.& $m_q$ & $m_{PS}$ & $m_V$ & $m_S$ & $m_{AV}$ \\
\hline
\hline
16x32 & 2.8 &.1455 & 3000& 0.69(2)&.0052(3)& .365(3)&  .429(3) & .369(4)  &  .434(4)\\
\hline
12x24 &2.6  &   .148 & 3000&0.83(1)& .0091(4) & .481(5) &.558(6)  &.491(5)  & .568(8)  \\
 \hline
8x16 & 2.3  &.1547&6000& 0.88(1)&-.009(6)& .680(5)&  .824(6)    & .709(5)    & .841(7)\\
\hline
\label{job parameter}
\end{tabular}
\end{table*}

\begin{figure*}
\includegraphics[width=7.5cm]{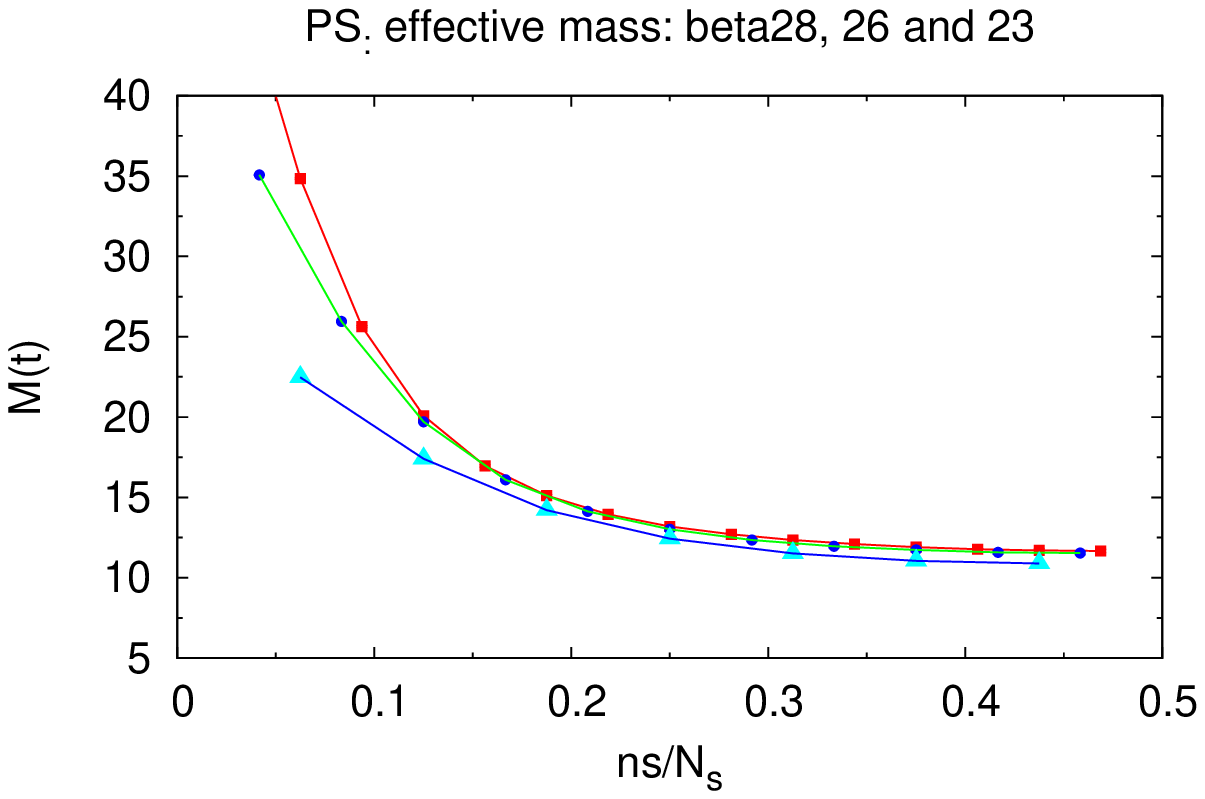}
\hspace{1cm}
\includegraphics[width=7.5cm]{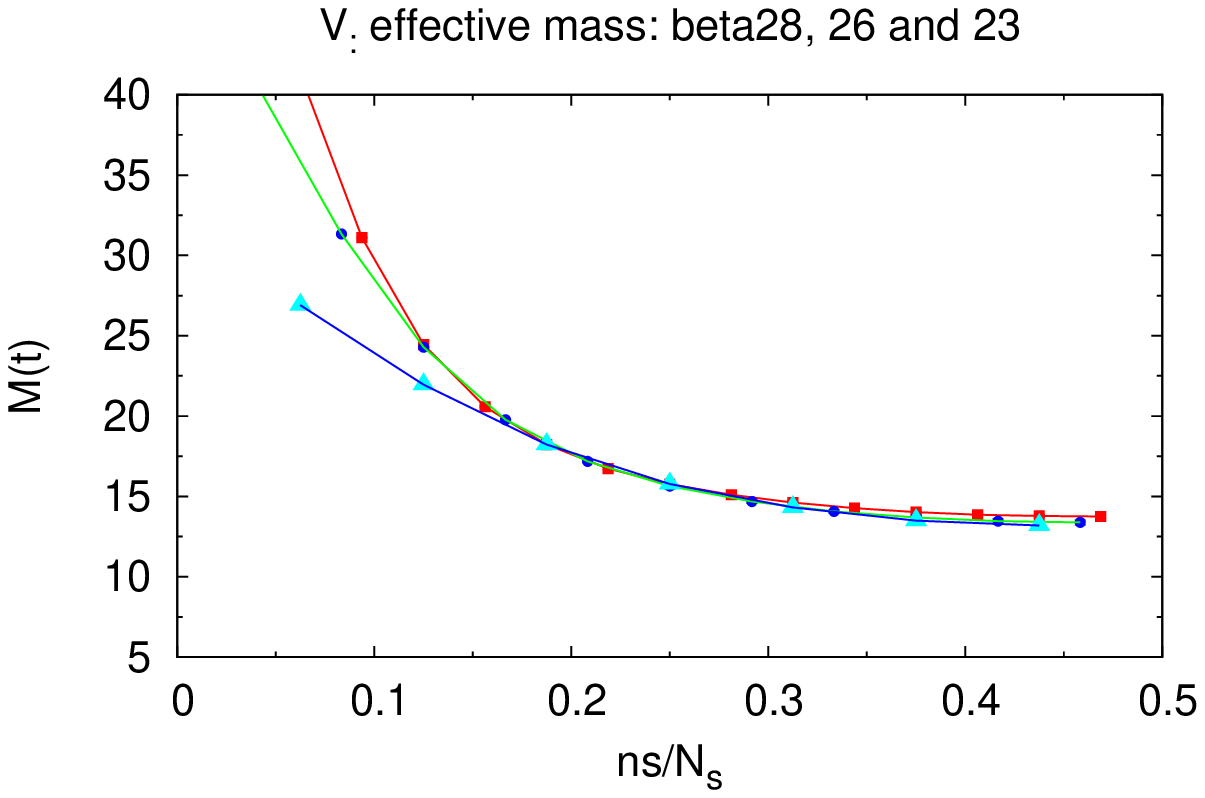}
\vspace{1cm}
\includegraphics[width=7.5cm]{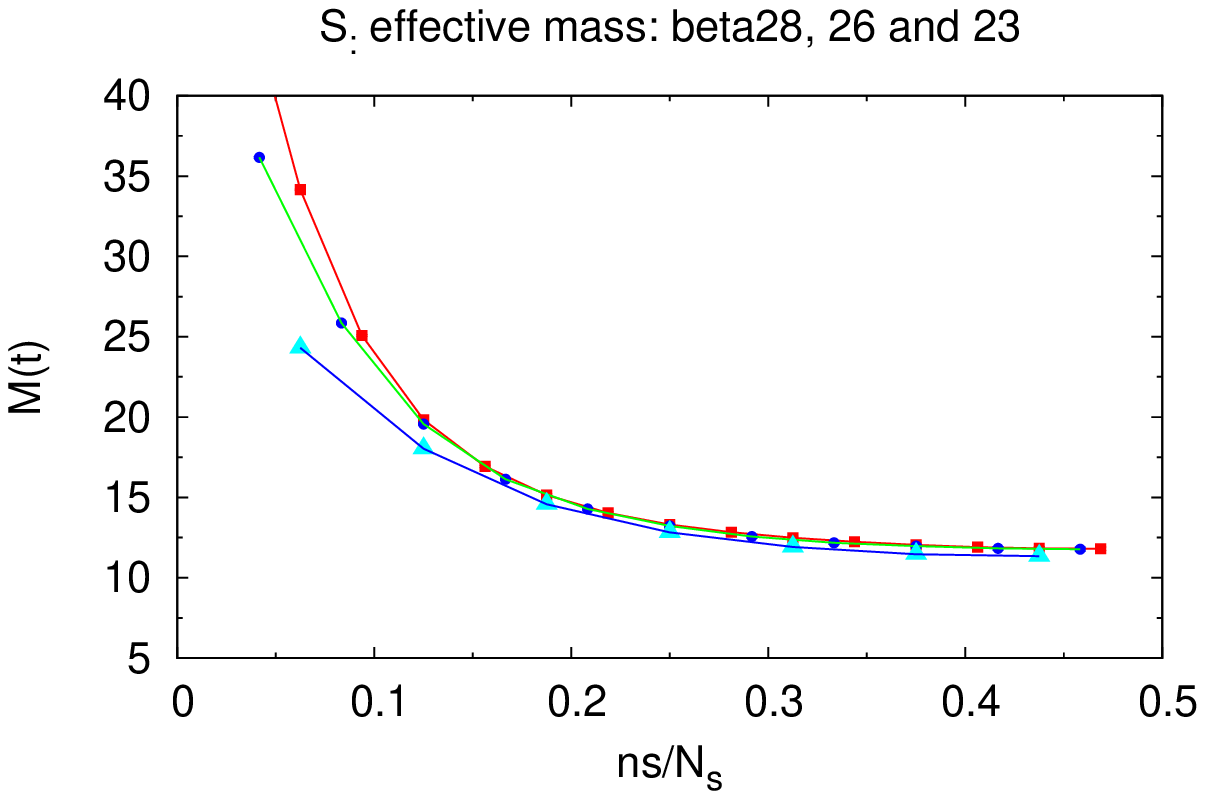}
\hspace{1cm}
\includegraphics[width=7.5cm]{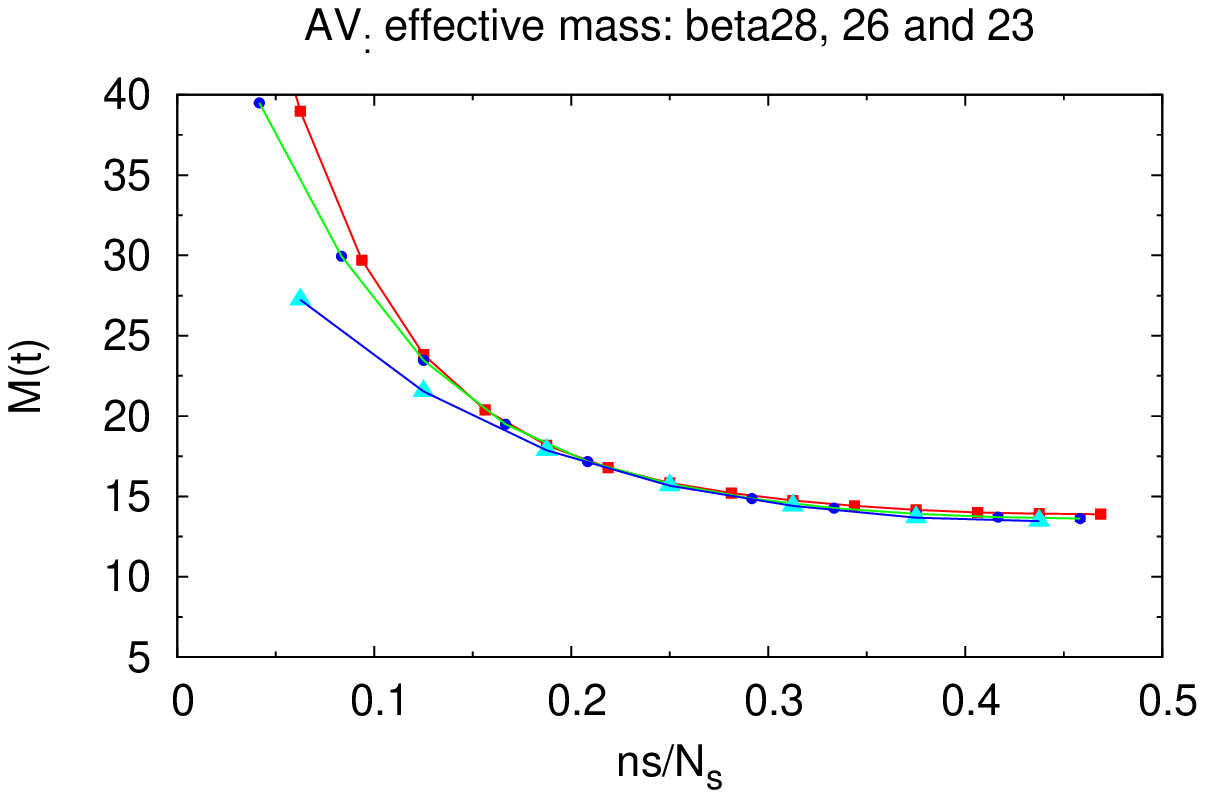}
\caption{(color online)  The effective spatial masses on the $16^3\times 8$ (blue), $24^3\times 12$ (green) and $32^3\times 16$ lattices (red) are overlaid: (left top) the pseudo-scalar meson; (right top) the vector meson.; (left bottom) the scalar meson.; (right bottom) the axial-vector meson. 
Lines connecting data are for guide of eyes.}  
\label{s-scaling}
\end{figure*}

\begin{figure*}
\includegraphics[width=7.5cm]{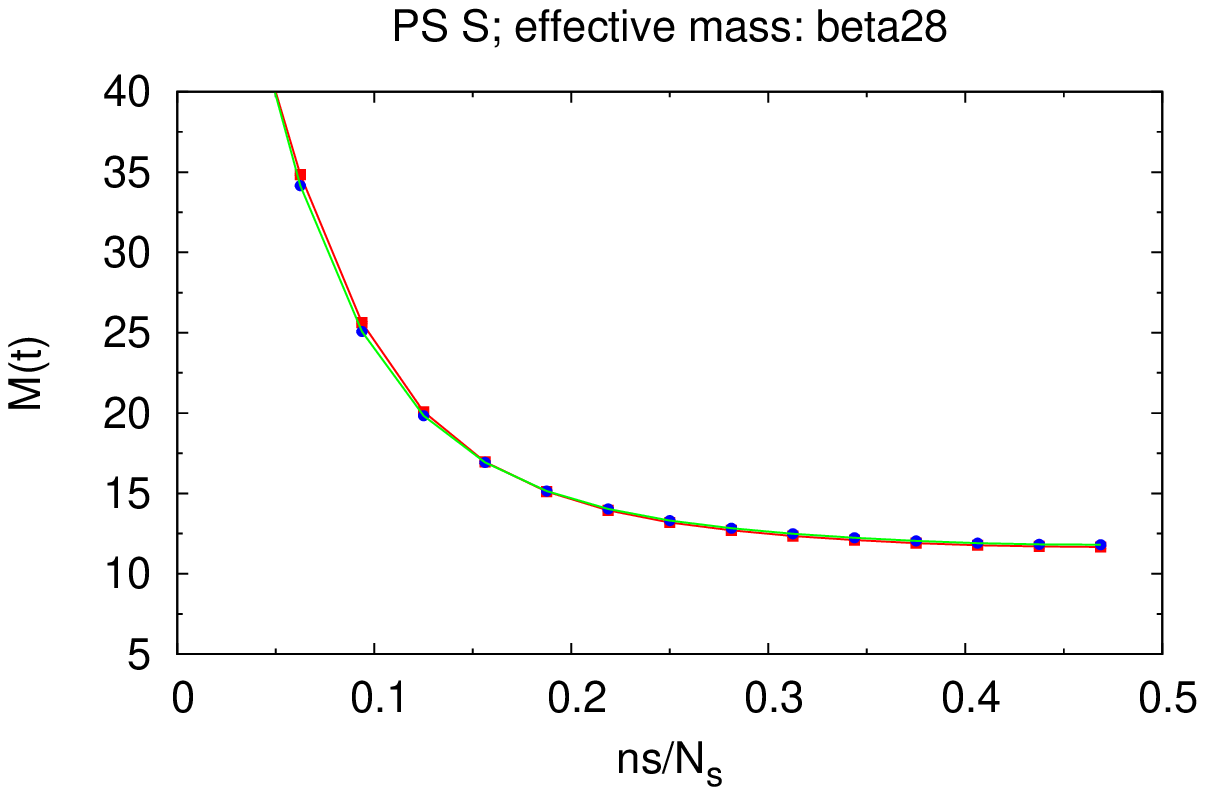}
\hspace{1cm}
\includegraphics[width=7.5cm]{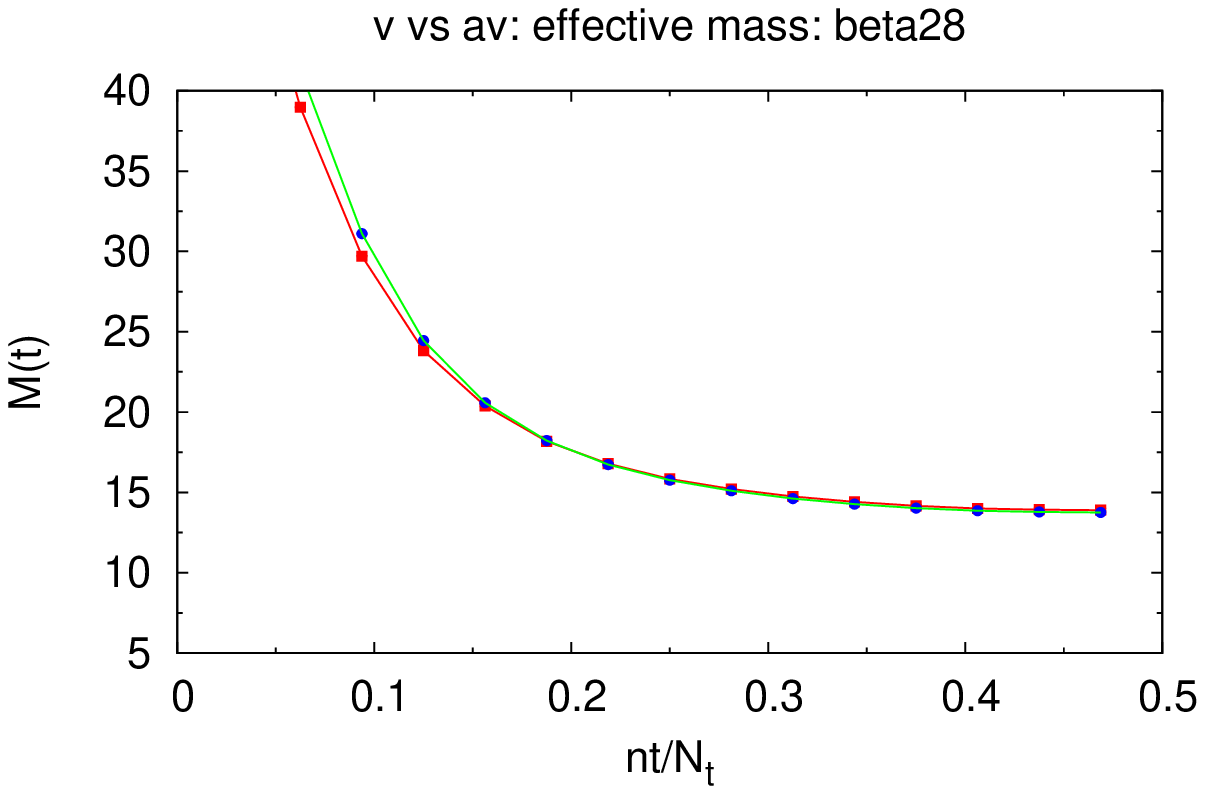}
\vspace{1cm}
\includegraphics[width=7.5cm]{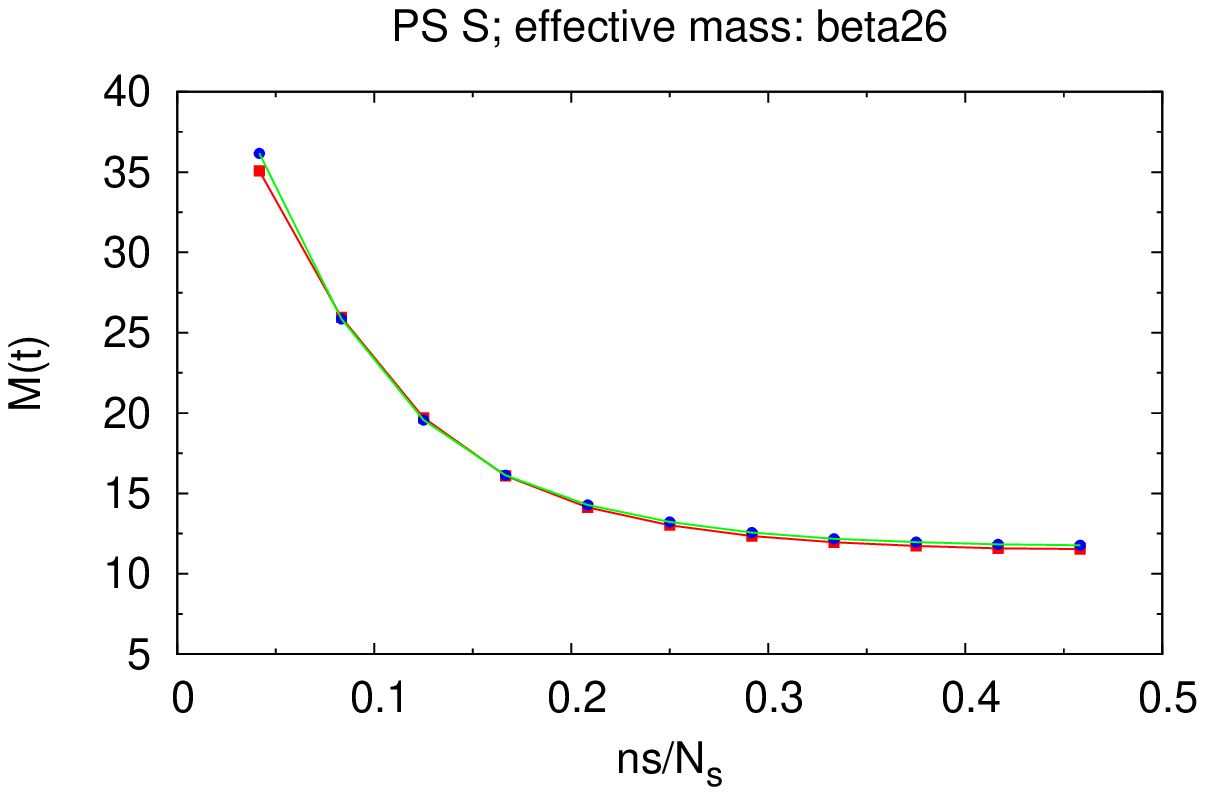}
\hspace{1cm}
\includegraphics[width=7.5cm]{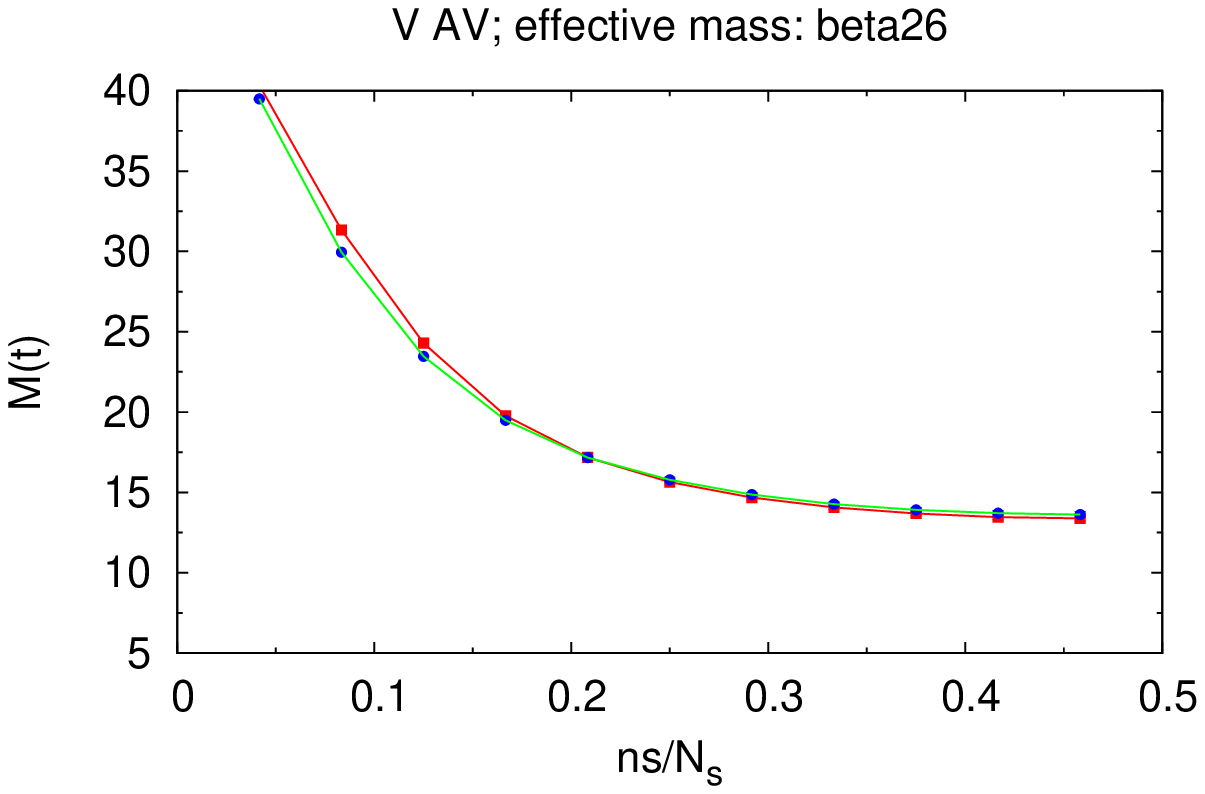}
\vspace{1cm}
\includegraphics[width=7.5cm]{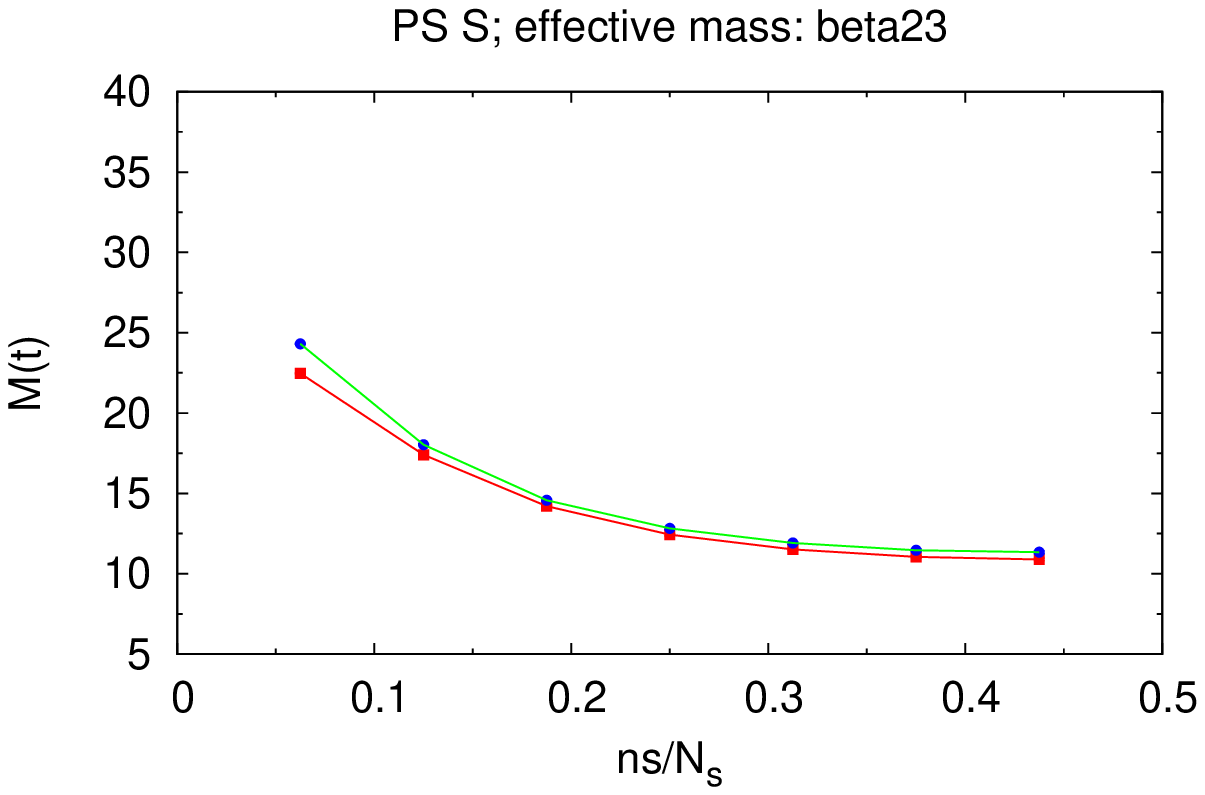}
\hspace{1cm}
\includegraphics[width=7.5cm]{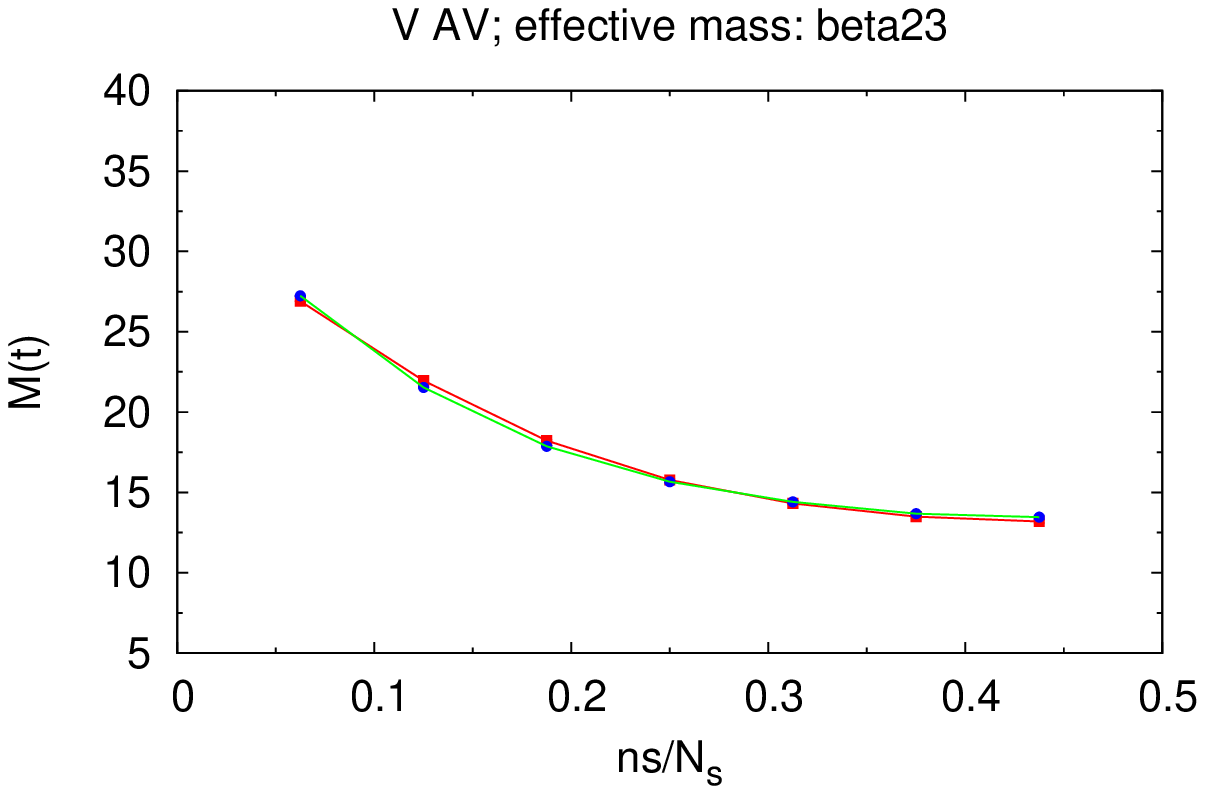}
\caption{(color online)  The effective spatial masses of the axial partners
 on the  $32^3\times 16,$ $24^3\times 12$ and   $16^3\times 8$ lattices :
(left) the $U_A(1)$ partners; the pseudo-scalar meson (green) vs the scalar meson (red):   (right) $SU_A(2)$ chiral  partners; the vector meson (green) vs the axial-vector meson (red).
On each size of lattices the parters are overlaid:  (top) $32^3\times 16$ lattice (middle) $24^3\times12$; (bottom) $16^3 
\times 8$ lattice.  Lines connecting data are for guide of eyes.}  
\label{s-scaling}
\end{figure*}


\begin{figure*}
\includegraphics[width=7.5cm]{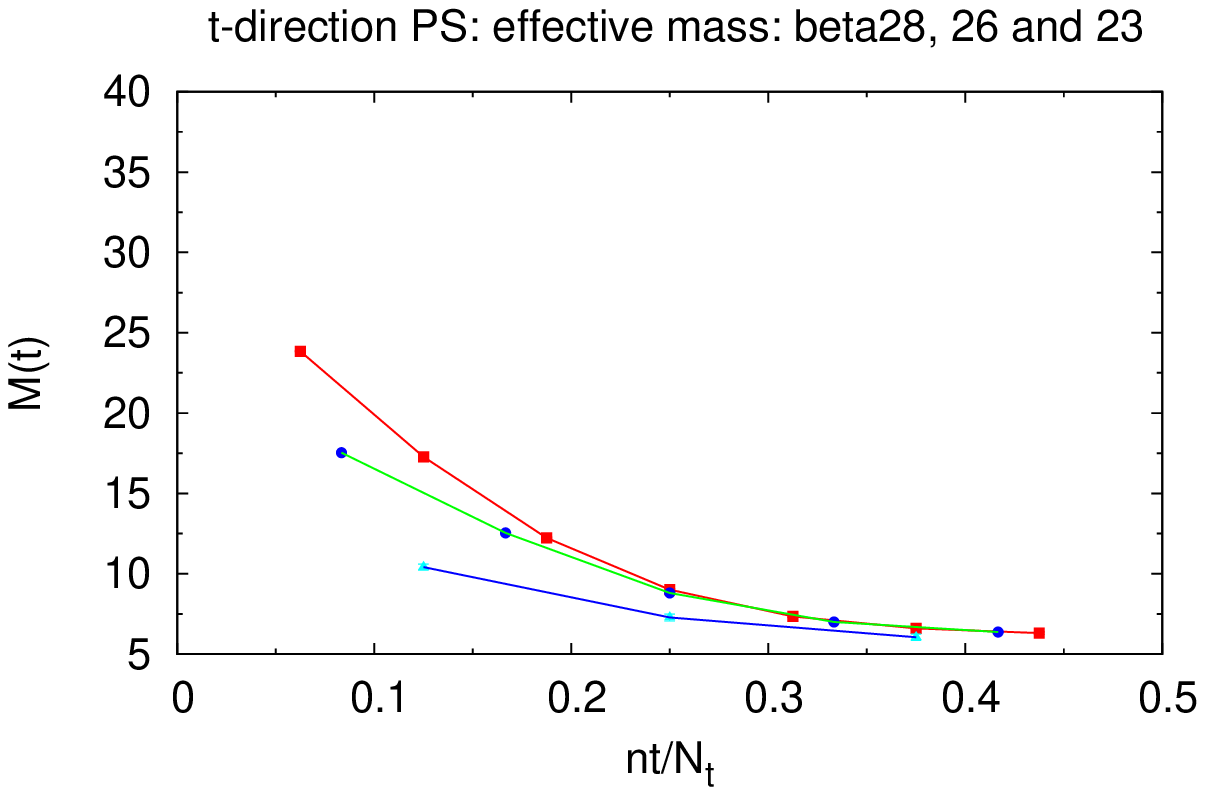}
\hspace{1cm}
\includegraphics[width=7.5cm]{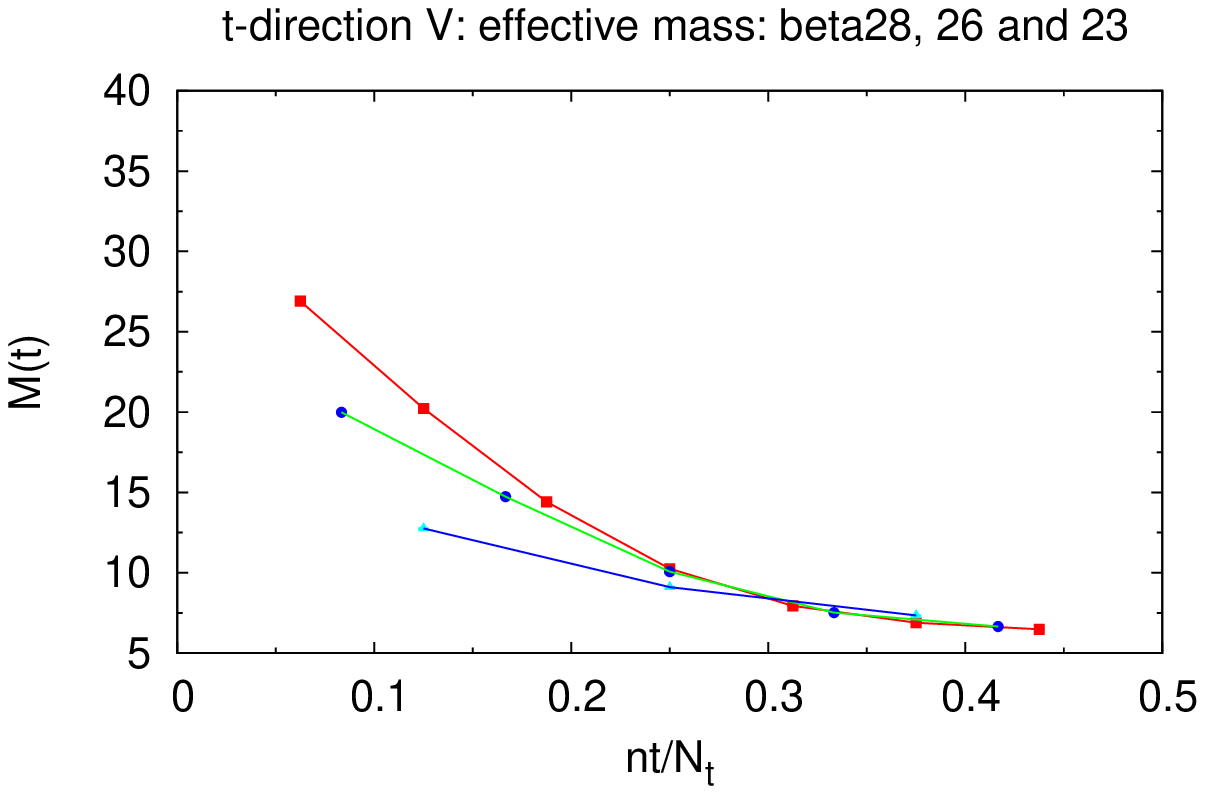}
\vspace{1cm}
\includegraphics[width=7.5cm]{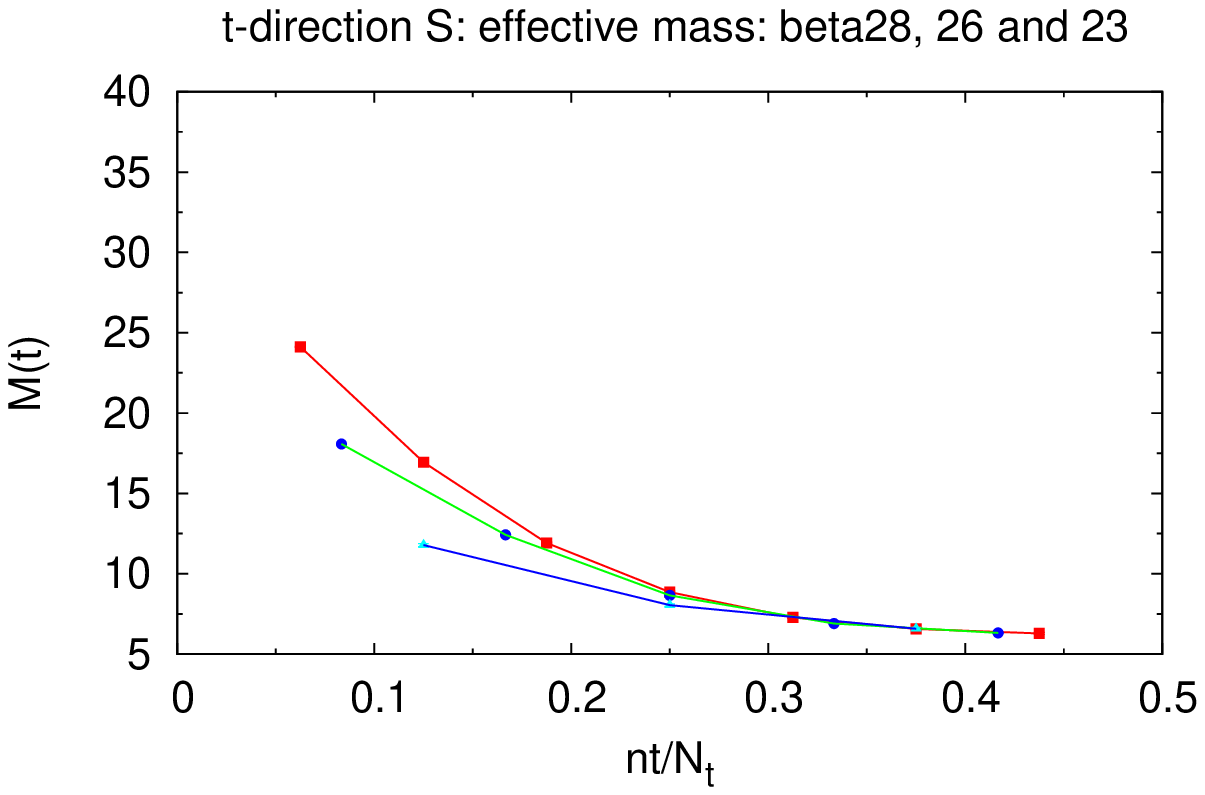}
\hspace{1cm}
\includegraphics[width=7.5cm]{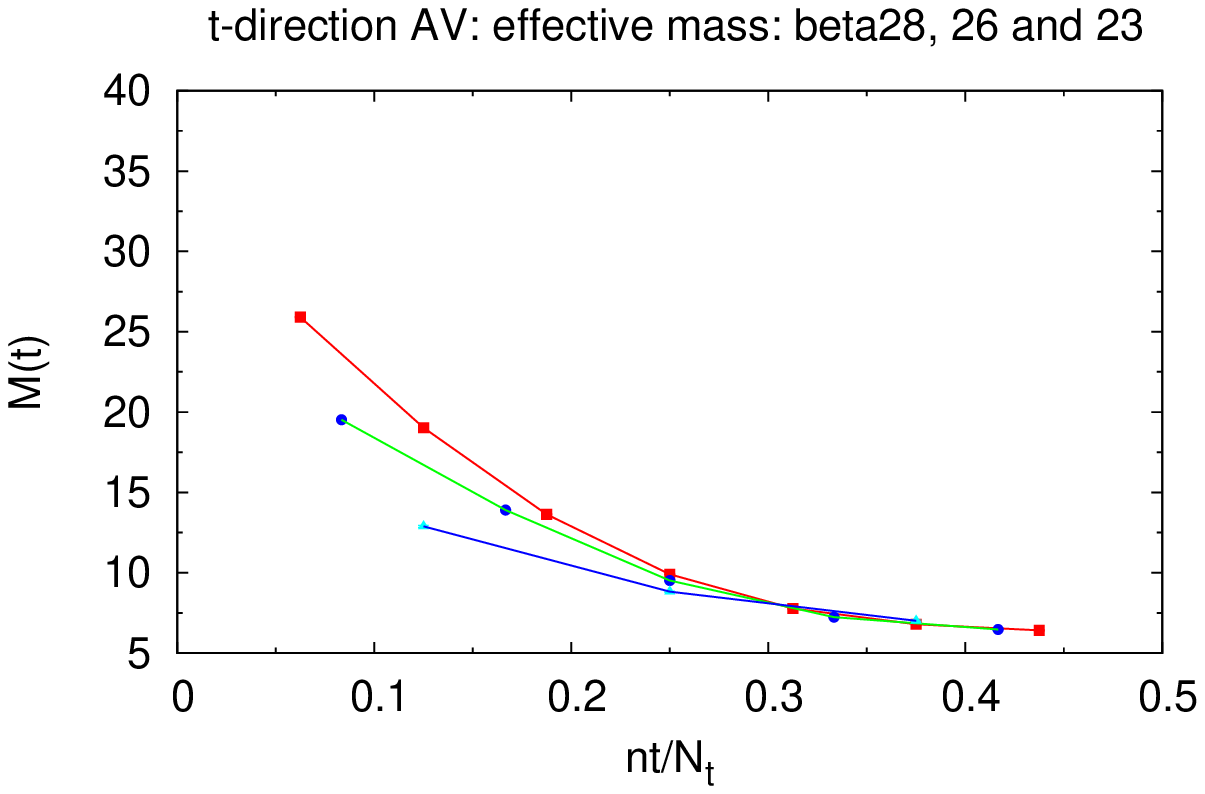}
\caption{(color online) 
 The effective temporal masses on the $16^3\times 8$ (blue), $24^3\times 12$ (green) and $32^3\times 16$ lattices (red) are overlaid: (left top) the pseudo-scalar meson; (right top) the vector meson.; (left bottom) the scalar meson.; (right bottom) the axial-vector meson. 
Lines connecting data are for guide of eyes.}  
\label{s-scaling}
\end{figure*}

\begin{figure*}
\includegraphics[width=7.5cm]{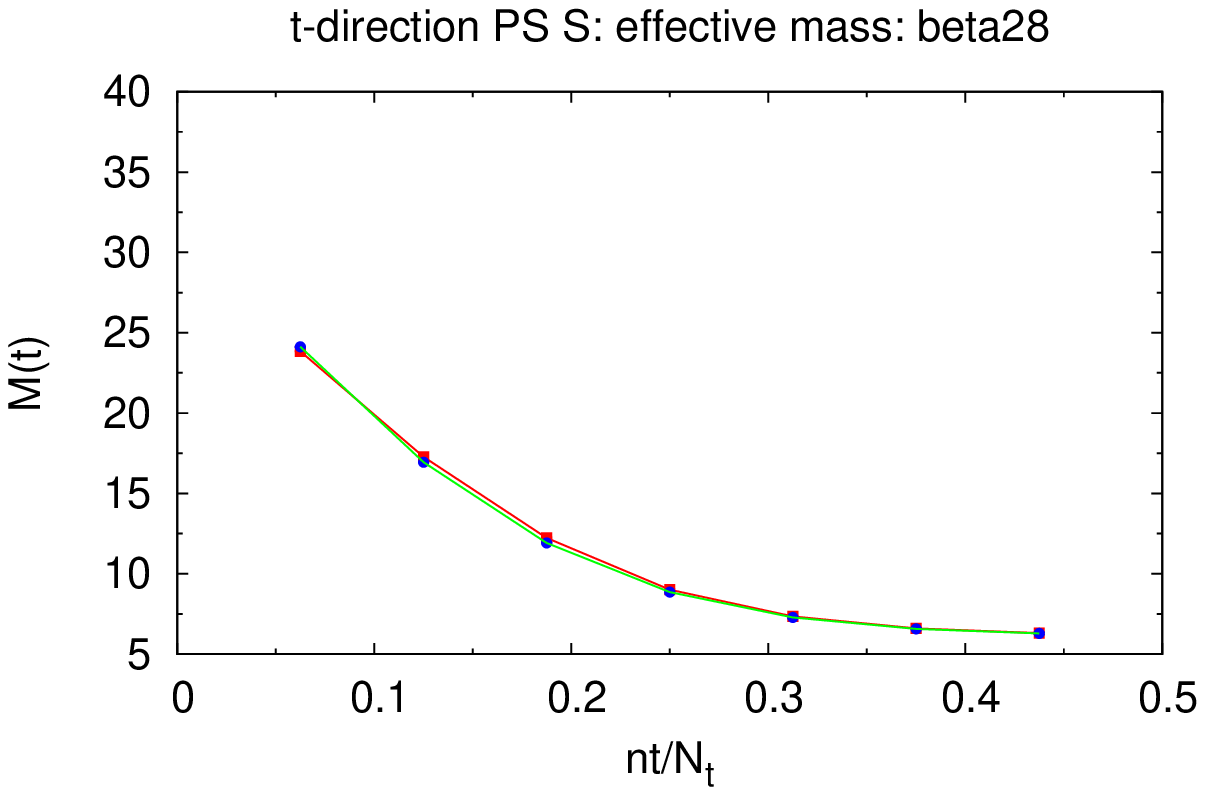}
\hspace{1cm}
\includegraphics[width=7.5cm]{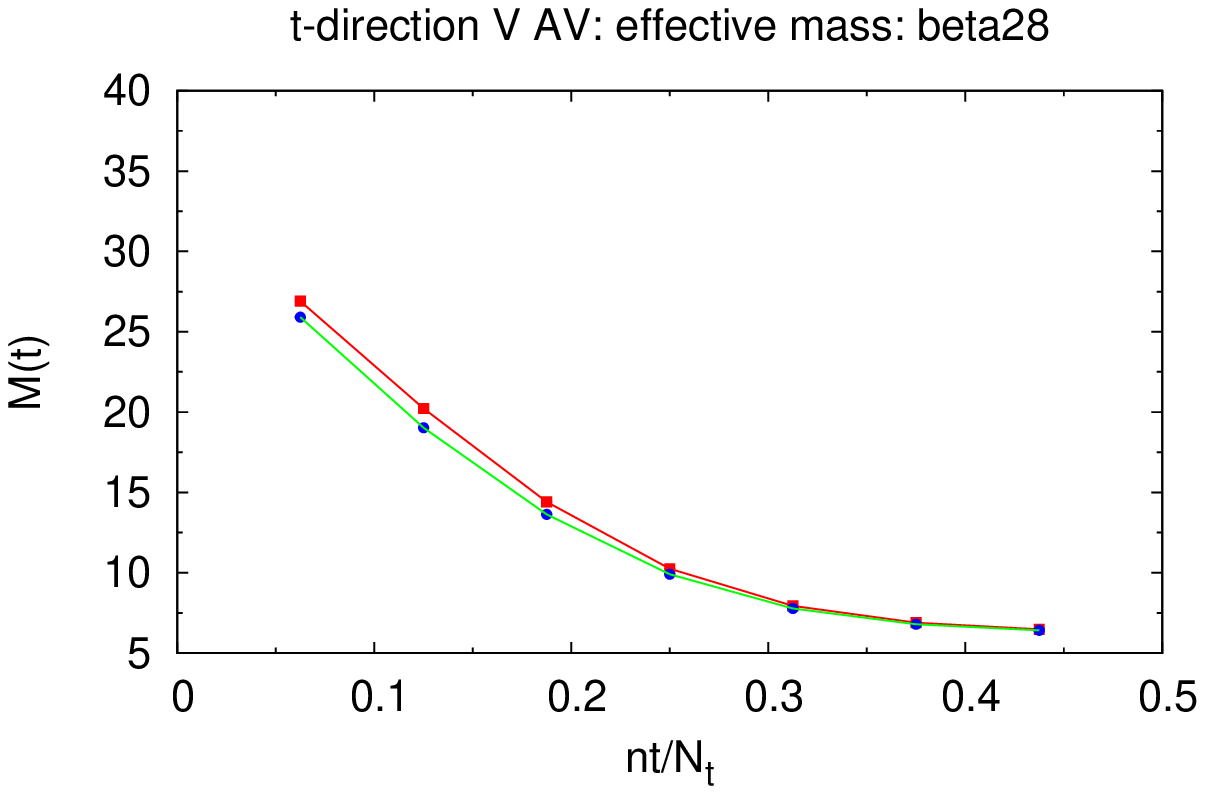}
\vspace{1cm}
\includegraphics[width=7.5cm]{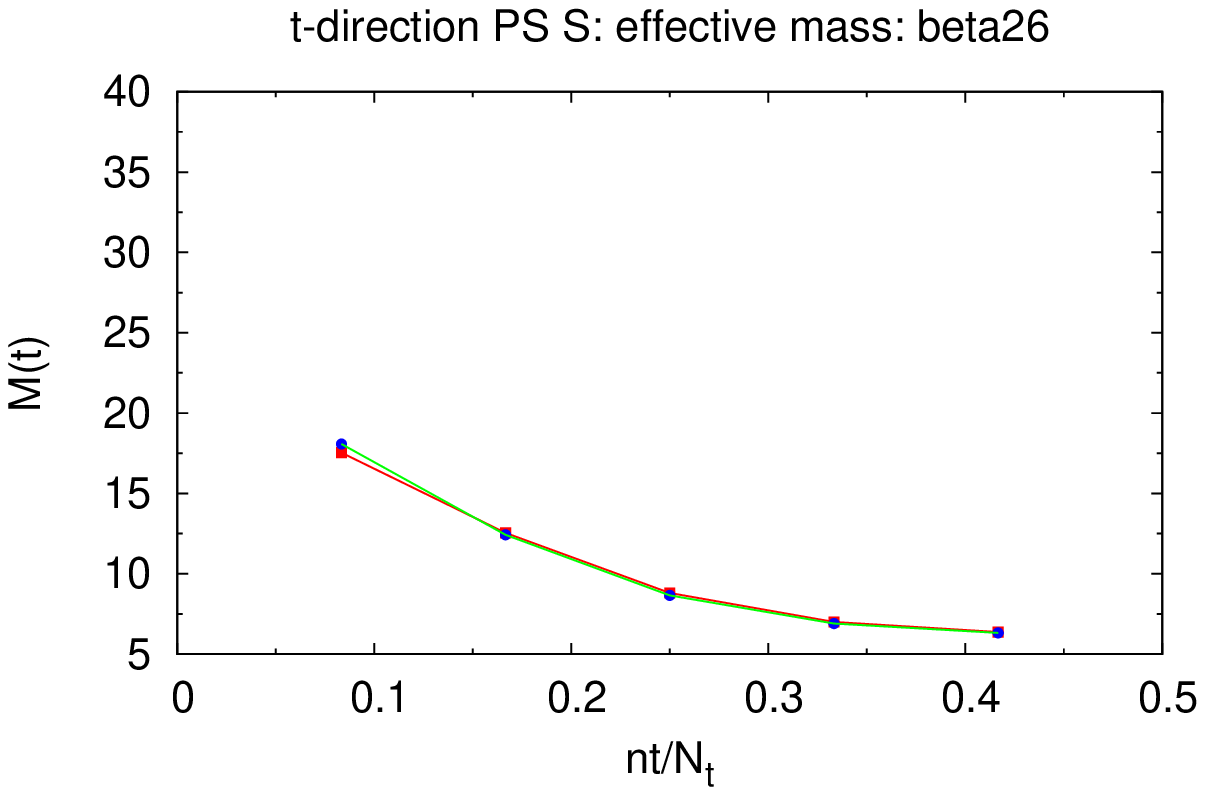}
\hspace{1cm}
\includegraphics[width=7.5cm]{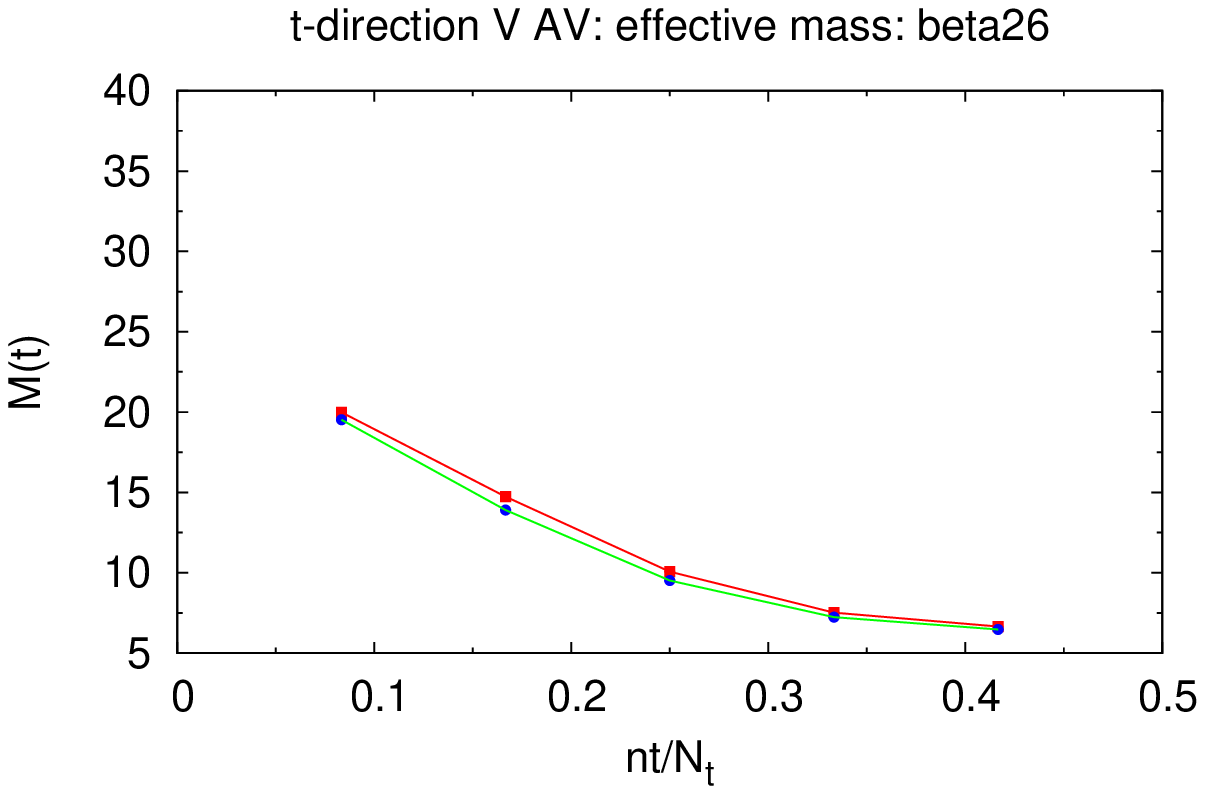}
\vspace{1cm}
\includegraphics[width=7.5cm]{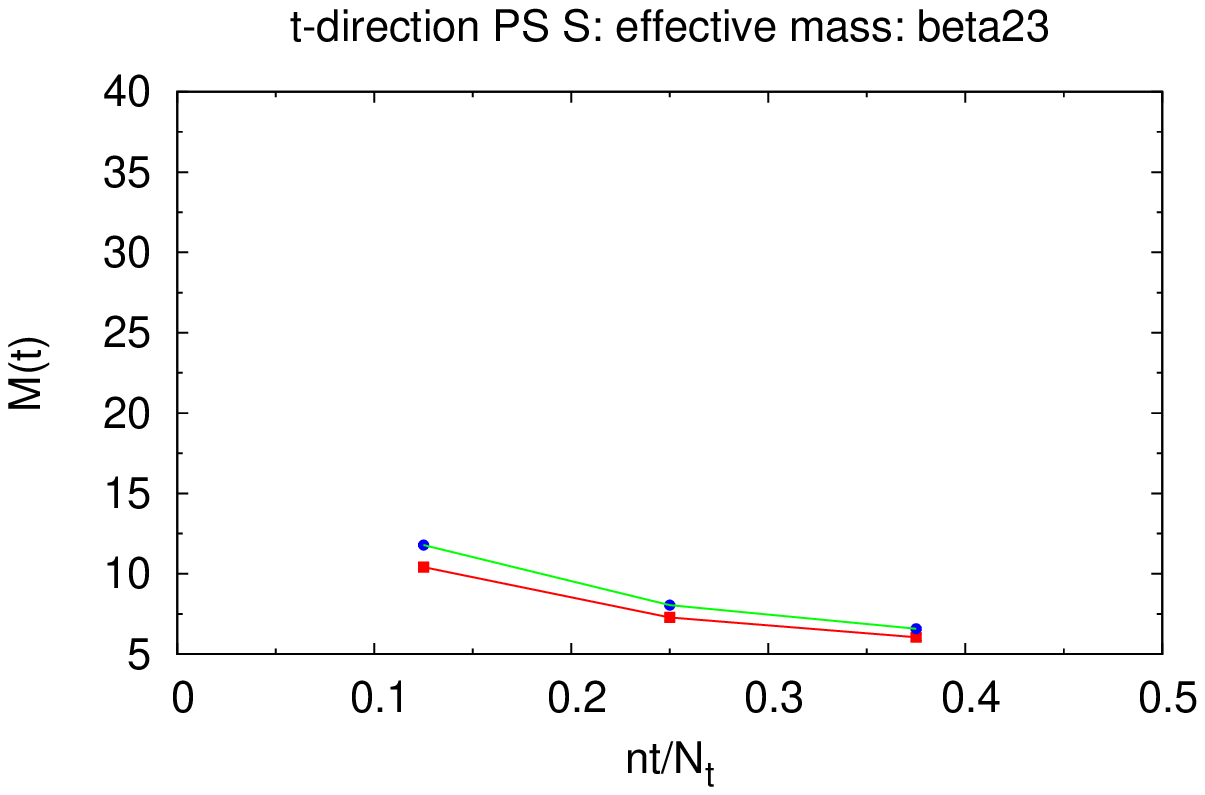}
\hspace{1cm}
\includegraphics[width=7.5cm]{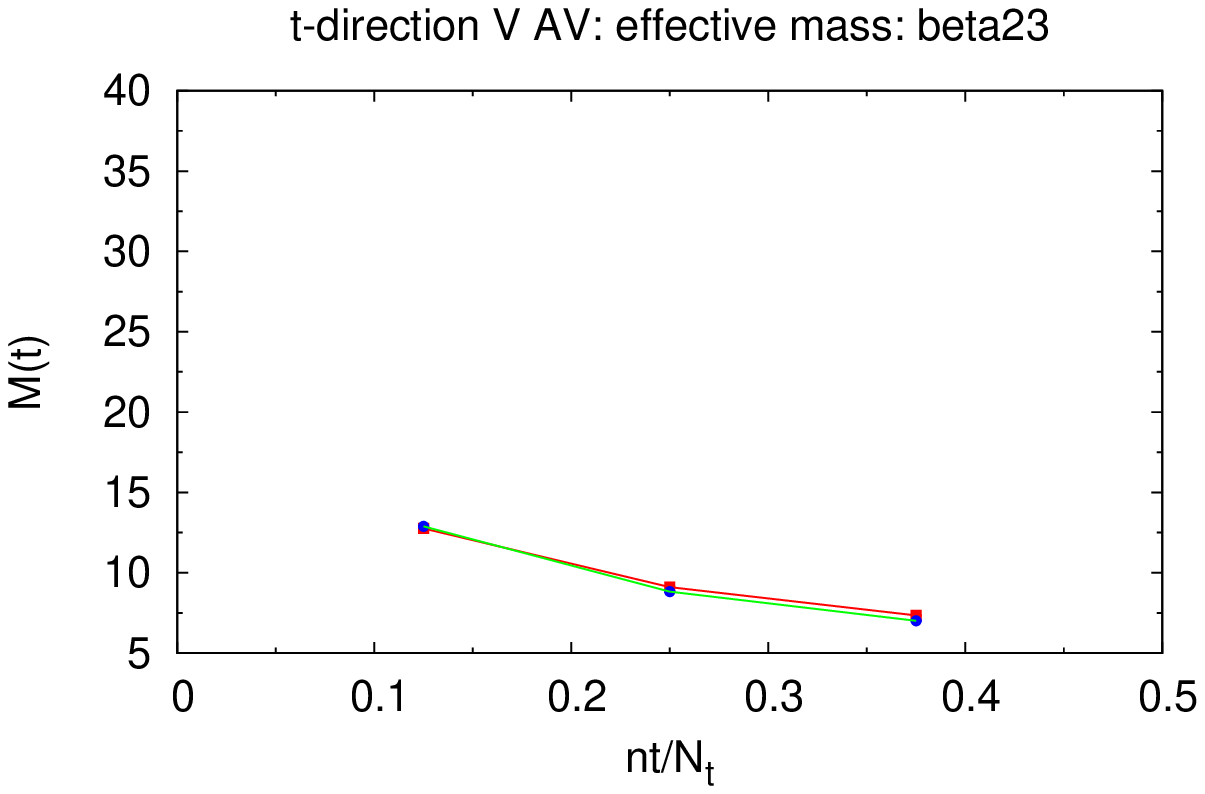}
\caption{(color online) 
 The effective temporal masses of the axial partners
 on the  $32^3\times 16,$ $24^3\times 12$ and   $16^3\times 8$ lattices :
(left) the $U_A(1)$ partners; the pseudo-scalar meson (green) vs the scalar meson (red):   (right) $SU_A(2)$ chiral  partners; the vector meson (green) vs the axial-vector meson (red).
On each size of lattices the parters are overlaid:  (top) $32^3\times 16$ lattice (middle) $24^3\times12$; (bottom) $16^3 
\times 8$ lattice.  Lines connecting data are for guide of eyes.}  
\label{s-scaling}
\end{figure*}


\begin{thebibliography}{0}
\providecommand{\natexlab}[1]{#1}
\providecommand{\url}[1]{\texttt{#1}}
\expandafter\ifx\csname urlstyle\endcsname\relax
  \providecommand{\doi}[1]{doi: #1}\else
  \providecommand{\doi}{doi: \begingroup \urlstyle{rm}\Url}\fi

\end{thebibliography}


\begin{thebibliography}{99}

\bibitem{Pisarski:1983ms} 
R.~D.~Pisarski and F.~Wilczek,
Phys.\ Rev.\ D {\bf 29}, 338 (1984).

\bibitem{Iwasaki:1996ya} 
  Y.~Iwasaki, K.~Kanaya, S.~Kaya and T.~Yoshie,
  Phys.\ Rev.\ Lett.\  {\bf 78}, 179 (1997).
 
 
\bibitem{AliKhan:2001ek} 
  A.~Ali Khan {\it et al.} [CP-PACS Collaboration],
  Phys.\ Rev.\ D {\bf 64}, 074510 (2001).
  
  
\bibitem{Ejiri:2009ac} 
  S.~Ejiri {\it et al.},
  Phys.\ Rev.\ D {\bf 80}, 094505 (2009).
 
   
  
\bibitem{Bazavov:2012qja} 
  A.~Bazavov {\it et al.} [HotQCD Collaboration],
  Phys.\ Rev.\ D {\bf 86}, 094503 (2012).
  
   
  
\bibitem{Aoki:2012yj} 
  S.~Aoki, H.~Fukaya and Y.~Taniguchi,
  Phys.\ Rev.\ D {\bf 86}, 114512 (2012).
  

\bibitem{Buchoff:2013nra} 
  M.~I.~Buchoff {\it et al.},
  Phys.\ Rev.\ D {\bf 89}, no. 5, 054514 (2014).

\bibitem{Pelissetto:2013hqa} 
  A.~Pelissetto and E.~Vicari,
  Phys.\ Rev.\ D {\bf 88}, no. 10, 105018 (2013).


\bibitem{Nakayama:2014sba} 
  Y.~Nakayama and T.~Ohtsuki,
  Phys.\ Rev.\ D {\bf 91}, no. 2, 021901 (2015).
  
  \bibitem{Nakayama:2016jhq} 
  Y.~Nakayama and T.~Ohtsuki,
  Phys.\ Rev.\ Lett.\  {\bf 117}, no. 13, 131601 (2016).
  
\bibitem{Sato:2014axa} 
  T.~Sato and N.~Yamada,
  Phys.\ Rev.\ D {\bf 91}, no. 3, 034025 (2015).
  
  
  \bibitem{Kanazawa:2015xna} 
  T.~Kanazawa and N.~Yamamoto,
  JHEP {\bf 1601}, 141 (2016).


\bibitem{Brandt:2016daq} 
  B.~B.~Brandt, A.~Francis, H.~B.~Meyer, O.~Philipsen, D.~Robaina and H.~Wittig,
  JHEP {\bf 1612}, 158 (2016).
  
\bibitem{Tomiya:2016jwr} 
  A.~Tomiya, G.~Cossu, S.~Aoki, H.~Fukaya, S.~Hashimoto, T.~Kaneko and J.~Noaki,
  arXiv:1612.01908 [hep-lat].


\bibitem{Ishikawa:2017hka} 
  K.-I.~Ishikawa, Y.~Iwasaki, Y.~Nakayama and T.~Yoshie,
  arXiv:1704.03134 [hep-lat].

 
 \bibitem{RG-improved}
Y. Iwasaki, UTHEP-118(1983); arXiv:1111.7059.  

   
\bibitem{Itoh:1984pr} 
  S.~Itoh, Y.~Iwasaki and T.~Yoshie,
  Phys.\ Lett.\  {\bf 147B}, 141 (1984).


  \bibitem{Bochicchio:1985xa} 
  M.~Bochicchio, L.~Maiani, G.~Martinelli, G.~C.~Rossi and M.~Testa,
  Nucl.\ Phys.\ B {\bf 262}, 331 (1985).

    
\bibitem{DelDebbio:2010ze} 
  L.~Del Debbio and R.~Zwicky,
  Phys.\ Rev.\ D {\bf 82}, 014502 (2010).


       
\bibitem{DeGrand:2009mt} 
  T.~DeGrand and A.~Hasenfratz,
  Phys.\ Rev.\ D {\bf 80}, 034506 (2009).



\bibitem{DelDebbio:2010hu} 
  L.~Del Debbio, B.~Lucini, A.~Patella, C.~Pica and A.~Rago,
  Phys.\ Rev.\ D {\bf 82}, 014509 (2010).
    

\bibitem{Hayakawa:2010gm} 
  M.~Hayakawa, K.-I.~Ishikawa, Y.~Osaki, S.~Takeda, S.~Uno and N.~Yamada,
  PoS LATTICE {\bf 2010}, 325 (2010).


\bibitem{Ishikawa:2013tua} 
  K.-I.~Ishikawa, Y.~Iwasaki, Y.~Nakayama and T.~Yoshie,
  Phys.\ Rev.\ D {\bf 89}, no. 11, 114503 (2014).

\bibitem{Ishikawa:2015nox} 
  K.-I.~Ishikawa, Y.~Iwasaki, Y.~Nakayama and T.~Yoshie,
  Mod.\ Phys.\ Lett.\ A {\bf 31}, no. 25, 1650150 (2016).


\bibitem{Cohen:1996ng} 
  T.~D.~Cohen,
  Phys.\ Rev.\ D {\bf 54}, R1867 (1996).

\bibitem{Lee:1996zy} 
S.~H.~Lee and T.~Hatsuda,
Phys.\ Rev.\ D {\bf 54}, R1871 (1996).

\bibitem{Evans:1996wf} 
  N.~J.~Evans, S.~D.~H.~Hsu and M.~Schwetz,
  Phys.\ Lett.\ B {\bf 375}, 262 (1996).

 
   
  \end{thebibliography}
\end{document}